\documentclass[preprint,aps,showpacs]{revtex4}
\usepackage{dcolumn}
\usepackage{bm}

\begin{document}
\preprint{}

\voffset 0.5in

\title{Quantum Transport Calculations Using Periodic Boundary Conditions}

\author{Lin-Wang Wang}
\affiliation{Computational Research Division,
Lawrence Berkeley National Laboratory, Berkeley, CA 94720
}


\date{\today}


\begin{abstract}
An efficient new method is presented to calculate the quantum transports using 
periodic boundary  conditions. This method allows the use of conventional
ground state ab initio programs without big changes. The computational 
effort is only a few times of a normal ground state calculation, thus it makes 
accurate quantum transport calculations for large systems possible. 
\end{abstract}


\maketitle

Quantum transport for molecules, nanowires, and nanodevices
is a fast growing research area in both experiment and theory, with 
the potential of replacing the current Si based technology after the Moor's
law reaches its limit in about 15 years. In the theoretical ballistic 
transport calculations,
a key step is to calculate the current 
via the Landauer formula:

\begin{equation}
I= {2e\over h} \int_{\mu_L}^{\mu_R} \sum_n T_n(E) dE,
\end{equation}
where $\mu_L$ and $\mu_R$ are left and right electrode Fermi energies
(assuming the current flows from right to left in z direction), 
and  $T_n(E)$ is the transmission coefficient for the $n$th right hand electrode channel (band) 
at energy E. There are two major ways to calculate $T_n(E)$. One is to use
the Green's function $G({\bf r},{\bf r'},E)$ of the system. However since $G$ is a 
double variable function, computationally this approach 
can be quite expensive, thus it is mostly used for localized basis set methods
\cite{Greens}. The other way to calculate $T_n(E)$ is to solve the following scattering
states:

\begin{equation}
H\psi_{sc}(r)=E\psi_{sc}(r)  
\end{equation}
and for $z\rightarrow \infty (-\infty)$:

\begin{equation}
\psi_{sc}(r)=\sum_n [A_n^{R(L)} \phi_n^{R(L)*}(r)+ B_n^{R(L)} \phi_n^{R(L)}(r) ],
\end{equation}
with conditions:   
$A_n^R=0$ except $A_m^R=1$ for one $m$, and $B_n^L=0$ (assuming ${dE_n^{R(L)}(k)/dk}>0$). 
In above, $H$ is the single particle Hamiltonian, and 
 $\phi_n^{R(L)}(r)=u_{n,k_n}(r) exp(i k_n^{R(L)}z)$ are right going running waves 
in the the right(R) and left(L) electrodes, and $\phi_n^{R(L)*}$ are the left going 
running waves.    
$E_n^{R(L)}(k_n^{R(L)})=E$ are the electrode band structure. 
The transmission coefficient
for channel m can be calculated as 
$T_m(E)=[\sum_n |A_n^L|^2 (dE_n^L(k)/dk)]/(dE_m^R(k)/dk)$, 
and the reflection coefficient can be calculated as
$R_m(E)=[\sum_n |B_n^R|^2 (dE_n^R(k)/dk)]/(dE_m^R(k)/dk)$. 
Transfer matrix method \cite{transmatrix1,transmatrix2}
and the Lippmann-Schwinger equation \cite{DiVentra} have been used to solve Eqs(2),(3). 
Unfortunately,  the transfer matrix 
method is rather complicated and computationally expensive to deal with the 
nonlocal pseudopotentials 
\cite{transmatrix2} and it is often plagued by the numerical instability due to the 
evanescent states in a multi-channel electrode \cite{transmatrix3}. 
On the other hand, 
the use of Lippmann-Schwinger equation \cite{DiVentra} requires the solution of a 
linear equation of the dimension of the full system, and it also needs the Green's
function of the two electrode system under a potential bias. As a result, currently 
this approach is only used for jellium electrode model and relatively small systems. 
Overall, compared to the more matured ground state calculations,  all the current methods for transport
calculations are complicated and computationally expensive, and they can only be used to calculate
relatively small systems although there is a strong need to study the 
transports of large molecules and nanostructures.  
Here, we present a new and simple approach which makes the transport calculation similar to  
the ground state calculation. In this approach, conventional periodic 
supercell methods and a specially designed perturbative approach are  used to solve Eqs(2),(3). 
This allows us to use modern ab initio 
total energy programs without much change. The computational effort is similar to 
a normal ground state local density approximation (LDA) calculation, hence it opens the door
for quantum transport studies for large systems.

To demonstrate our method, we have chosen a benzene molecule connected by two Cu quantum 
wires as shown in Fig.1. This system is chosen since the conductivity of
the benzene molecule is well studied \cite{DiVentra,Reed} and similar quantum wires have been used as electrodes in 
previous quantum transport calculations \cite{Roland}. Two hydrogen atoms at the two ends of 
the benzene molecule are replaced by two sulfur atoms, which are bonded to two central
Cu atoms at the electrode. The atomic positions of the molecule are relaxed at the zero 
voltage bias under LDA calculation.   
We have used norm conserving 
pseudopotentials and 30 Ryd planewave cutoff with a standard
planewave LDA program \cite{PEtot}.  
We have included 5 and 6 unit cells in the left and right electrodes respectively, 
and they are connected at boundary B in Fig.1 by periodic boundary condition.  
The x, y dimensions of the supercell are 3 times the width of the Cu wire to avoid
possible neighbore-neighbore interactions. 
After the Kohn-Sham single particle potential $V_0(r)$ is obtained from a LDA selfconsistent 
calculation at the zero bias,  
we have added a 
potential $V/2 sin(\pi z/L')$ in the central region of the molecule and
shifted the rest of the right (left) electrode by $V/2$ 
($-V/2$) to get the potential $V_V(r)$ for a bias V system. 
Although a selfconsistent treatment can be achieved straight forwardly 
under the current approach 
[since the scattering states of Eq(2) will be calculated],
the current nonselfconsistent treatment for finite bias V is sufficient in illustrating the new
methodology. Note that, there is a jump of $V_V(r)$ at the boundary B, 
but that is not a problem in our numerical calculations.

Figure 2 shows the band structures $E_n(k_z)$ of the quantum wire electrode. There are 9 Cu atoms in 
each unit cell of the electrode. To simplify our calculation and analysis, we did not include the 3d 
electrons in our pseudopotential. Although this will introduce a significant error in the electrode 
total energy, the electronic structure near the Fermi surface and the related transport properties
are intact. However, for bias larger than 2 V, our electrode should be considered
as a model electrode due to the lack of Cu 3d electrons.

To solve the scattering states of Eqs(2),(3), we first calculate the eigenstates 
$\{ \psi_i(r)$, $E_i \}$ 
of the periodic supercell under $V_V(r)$ with a ${\bf K}_z$ point (e.g, ${\bf K}_z= \pi/2L_z$, where $L_z$ is the length of the
supercell and this ${\bf K}_z$ is not the $k_z$ of the electrode as in Fig.2) using our standard LDA program \cite{PEtot}. 
Let's first assume that, using some methods,  we can generate $l$ degenerated states $\psi_{i,(l)}(r)$ 
they all satisfy the Schrodinger's equation $H\psi_{i,(l)}=E_i \psi_{i,(l)}$ (however, for our purpose, 
this equation needs only to be satisfied within the interior of the supercell, not near the boundary B
of Fig.1). 
The idea is to use a linear combination of these states to construct the scattering states of Eq(2). 
First, within the R(L) electrode, $\psi_{i,(l)}(r)$ can be decomposed into the electrode states
$\phi_n^{R(L)}(r)$ just as in Eq(3). From a given $E_i$, the available $n$ and $k_n^{R(L)}$ can be found
from Fig.2, or say: 
$E_n(k_n^{R(L)})+\mu_{R(L)}=E_i$. The corresponding $\phi_n^{R(L)}$ is then generated by numerical  interpolations
from pre-calculated electrode states. 
The expansion coefficients $A_n^{R(L)}(i,l)$, $B_n^{R(L)}(i,l)$  for
wavefunction $\psi_{i,(l)}$ can be easily calculated from the functional products like: 
$\int_{\Omega} \psi_{i,(l)} \phi_n^* d^3r$ and $\int_{\Omega} \phi_m \phi_n^* d^3r$, where $\Omega$ 
is one electrode unit cell at the middle of the electrodes as shown in Fig.1. 
We found that,  after including the possible evanescent 
states \cite{footnote}, 
this expansion typically captures more than $99.99\%$ of the weight of the original 
$\psi_{i,(l)}$. The next step is to make a linear combination of $\psi_{i,(l)}$ to get the 
scattering state $\psi_{i,sc}$ of Eq(2):

\begin{eqnarray}
\psi_{i,sc}= & \sum_l C_l \psi_{i,(l)} \\ \nonumber
= & \sum_n \sum_l [C_l A_n^{R(L)}(i,l) \phi_n^*(r) +  C_l B_n^{R(L)}(i,l) \phi_n(r) ],  
\end{eqnarray}
here the second equation and the R and L are for $r$ within the right and left electrodes respectively. 
Note that, due to the use of supercell ${\bf K}_z$ point and both $\psi_i$ and $\psi_i^*$ are used as
$\psi_{i,(l)}$, $\psi_{i,sc}$ is no longer periodic
at the supercell boundary B.  To make $\psi_{i,sc}$ in Eq(4) as one scattering state of Eq(3), we need
to make it satisfy the boundary conditions of Eq(3):$A_{n\ne m}^R=0$, $B_n^L=0$, by selecting 
$C_l$. In order to have a solution for $C_l$, we need N independent $\psi_{i,(l)}$ ($l=1,N$), 
if there are N nonzero contributing electrode states $n$ in Eq(4) [counting
both the left and right electrodes, but $\phi_n$ and $\phi_n^*$
are counted as one]. 

Notice that, since both $\psi_i(r)$ and $\psi_i^*(r)$ satisfy the Schrodinger's equation
$H\psi=E_i\psi$,  for systems with only a single channel (N=2),
the eigen state $\psi_i(r)$ itself is enough to construct the scattering states $\psi_{i,sc}$ from 
Eq(4). So, the main task here is for multi-channel cases. In our example, from Fig.2 we see
that for a given energy $E_i$, we could have 4-5 channels. To get the N degenerated 
 $\psi_{i,(l)}$, we will use a
perturbative approach. First, we will construct $W_m(r)=u_{m,\Gamma}(r)$ (the $k=0$
$m$th electrode state) when $z$ is within the last unit cell
of the right electrode near boundary B of Fig.1, and $W_m(r)=0$ for all the other $z$ (see Fig.1).  
We will add $\beta |W_m><W_m|$ as a perturbation in the original $H$, and solve
the following eigenstate equation
using our standard LDA program (e.g, using conjugate gradient method):

\begin{equation}
 H \psi_{i,m}' + \beta <W_m|\psi_{i,m}'> W_m = (E_i + \Delta E_{i,m}) \psi_{i,m}'.
\end{equation}
Here $\beta$ is a very small number, hence $\Delta E_{i,m}$ and 
$\Delta \psi_{i,m}\equiv \psi_{i,m}'-\psi_i$ are both small. Suppose we have solved the above
equations for two different $m$'s: $m_1$ and $m_2$. Then we can construct 
$\psi_{i,(l)}=F_1 \Delta \psi_{i,m_1}+F_2 \Delta \psi_{i,m_2}$, with 
$F_1\Delta E_{i,m_1}+F_2\Delta E_{i,m_2}=0$. After dropping the second order terms
$\Delta E_{i,m_{1,2}} \Delta \psi_{i,m_{1,2}}$ we have:

\begin{equation}
 H \psi_{i,(l)} + \beta \sum_{j=1,2} F_j <W_{m_j}|\psi_{i,{m_j}}'> W_{m_j} 
= E_i \psi_{i,(l)}
\end{equation}

Notice that the $W_{m_j}$ terms are nonzero only near the boundary B, so for all the other
places, we have $H \psi_{i,(l)}= E_i \psi_{i,(l)}$.  Thus, $\psi_{i,(l)}$ are the wavefunctions we needed.  
In the simple cases, when there are N total electrode states $\phi_n^{R(L)}$ with nonzero 
components in the expansion of  
$\psi_i(r)$ in Eq(4), there will be N/2 right electrode states $\phi_n^R$. Then 
the perturbations by the related  N/2 $W_m$ states (which have the same characteristics
and cross section symmetries as $\phi_n^R$)
will introduce N/2 indepedent perturbative wavefunction changes $\Delta \psi_{i,m}$. 
These $\Delta \psi_{i,m}$ will generate  $(N/2-1)$
independent $\psi_{i,(l)}$ states (besides the original $\psi_i$). Thus, the total number of 
$\psi_{i,(l)}$ states (counting also $\psi_{i,(l)}^*$) is just N, the exact number we 
need to construct 
the scattering state $\psi_{i,sc}$ from Eq(4).  This
argument remains true when there are evanescent states or the number of electrode states
in the left and right electrodes are not the same. Thus, using this procedure, we are 
guaranteed that there will be enough $\psi_{i,(l)}$ states for a given $\psi_i$ to generate
a few corresponding scattering states $\psi_{i,sc}$.

From the supercell eigenstates $\{ \psi_i(r), E_i \}$, we
can generate a set of  $\{ k_n^R \}$ from $E_n(k_n^R)+\mu_R=E_i$. These $\{ k_n^R \}$ 
are shown in Fig.2 as the crosses for a 1V bias case 
(using all the $\psi_i(r)$ with $E_i$ between the two horizontal arrows
in Fig.2). 
As can be described by a phase accumulation model\cite{PAM},  on each band, these $k_n^R$ have roughly 
equal distances 
and their total number roughly equals the number of electrode unit cells.  
We have typically used 6 $\Gamma$ point electrode states as $W_m$ in Eq(5) starting from the lowest
band as annotated in Fig.2. 
This means we have to solve $\{ \psi_{i,m}' \}$ of Eq(5) 6 times using
$\{ \psi_i \}$ as the initial wavefunctions (Notice that, this number 6 is roughly the
number of channels in the problem. The same prefactor is needed in the calculations of other
methods like the transfer matrix or Lippmann-Schwinger equation). 
After $\{ \psi_{i,m}' \}$ are calculated, using Eq(4), 
we can construct a scattering state from each of these $k_n^R$ shown in Fig.2. 
Two of these constructed scattering states are shown in Fig.3. Notice that 
the dash lines are $\sum_l C_l \psi_{i,(l)}$ of Eq(4), while the solid lines 
are the electrode state decompositions [the second line of the Eq(4)]. 
Within the electrode, 
the electrode state decomposition gives a very accurate description of the
total wavefunction. From these scattering states, the transmission coefficients
$T_m(E_i)$ can be calculated, and are shown in Fig.4 as the symbols. 
The calculated $T_m(E_i)+R_m(E_i)$ is typically very 
close to 1, indicating the numerical stability of the current method.

Notice that, unlike the other  approaches discussed before where the scattering 
states of arbitrary energy E can be solved, here only the scattering states of
energy $\{ E_i \}$ are calculated. This translates into finite number of $\{ k_n^R \}$ 
points as shown in Fig.2 and Fig.4. In Eq(1), we need all the energies between 
$\mu_L$ and $\mu_R$. This is an complete analogy with the k-point integration problem
in conventional ground state bulk calculation [the energy integral in Eq(1) can also be changed 
into a k-point integral of the electrode band structure of Fig.2]. Thus, similar
to the conventional bulk calculations, 
here we will use an interpolation scheme to carry out the integral in Eq(1). First, 
if the number of $\{ k_n^R \}$ is not enough in Fig.4, we can choose a different supercell ${\bf K}_z$ 
in our supercell calculations (or  change the potential $V_V(r)$ near boundary B),
and repeat the above procedure. That will give us more
$\{ E_i\}$ and $\{ k_n^R \}$ points.  In our system, we find one ${\bf K}_z$ calculation is sufficient. 
We have used a smooth curve $f_n(k)$ to interpolate the $ln(T_n(k_n^R))$ points shown in Fig.4. 
More specifically, we have minimized: 
$\sum_{k_n^R} |ln(T_n(k_n^R))-f_n(k_n^R)|^2+ \gamma \int |d^2 f_n(k)/dk^2|^2 dk $, with $\gamma\sim 1$.  
Numerically, this corresponds to a simple linear equation with discretized $k$ points.  
The resulting curve is shown in Fig.4 for the 1V bias case. Using these $f_n(k)$, we can 
calculate the total transmission  $T(E)=\sum_n T_n(E)$ of the system. The results are shown 
in Fig.5 for different biases. We see that $T(E)$ is influenced strongly by two factors.  One
is the relative energy levels between the electrode states and the molecule states. When the 
bias increases, the molecular levels drop relative to the right electrode state levels. 
As a result, the magnitude of the transmission decreases near the region of  -3 eV. 
Another factor is the band structure of the electrode. There is a well shape of $T(E)$ near
-0.3 eV. This is caused by band gaps of the 2,3 bands at the $X'$ point in Fig.2. The $T(E)$ also shows
a big drop at -3.3 eV. This is due to the end of 2,3 bands at the $\Gamma$ point.

After $T(E)$'s of Fig.5 are obtained, a simple energy integration between -V to 0 will give us 
the total current $I$. The resulting $I(V)$, and the conductance $dI/dV$ are shown in Fig.6.
We do see the well known peak and dip for this system in the conductance around 2V. Our calculated
peak and dip positions of 1.8V and 2.3V corresponds well with the experimental results 
of 1.4V and 2.4V \cite{Reed}, and this agreement is better than previously calculated 
results \cite{DiVentra}.  We also see the marks of the electrode electronic
structures. Near 0.3V, the conductance shows a well shape, again due to the band gaps
of 2,3 bands at the $X'$ point. Above 3.3V, we see a big drop, then the negative conductance. 
The drop is due to the end of the 2,3 bands at the $\Gamma$ points, and the negative 
conductance is because the conducting electrode levels (energy window) are moving away from 
the conducting molecular levels. Here we see that, the electronic structure of the electrode 
is extremely important in determining the overall conductance of the system.

In summary, we have presented a simple and numerically stable scheme to calculate the quantum transport. Within 
this scheme, the conventional periodic supercells and ground state ab initio programs can be
used without much change. The computational effort is only a few times of a conventional ground state 
calculation. This promised quantum transport calculations for much larger systems which cannot be 
tackled by othe methods. The implementation of this method is simple and straight forward based on any 
conventional ground state ab initio programs. 

This work was supported by U.S. Department
of Energy under Contract No. DE-AC03-76SF00098. This research
used the resources of the National Energy Research Scientific
Computing Center.


\newpage

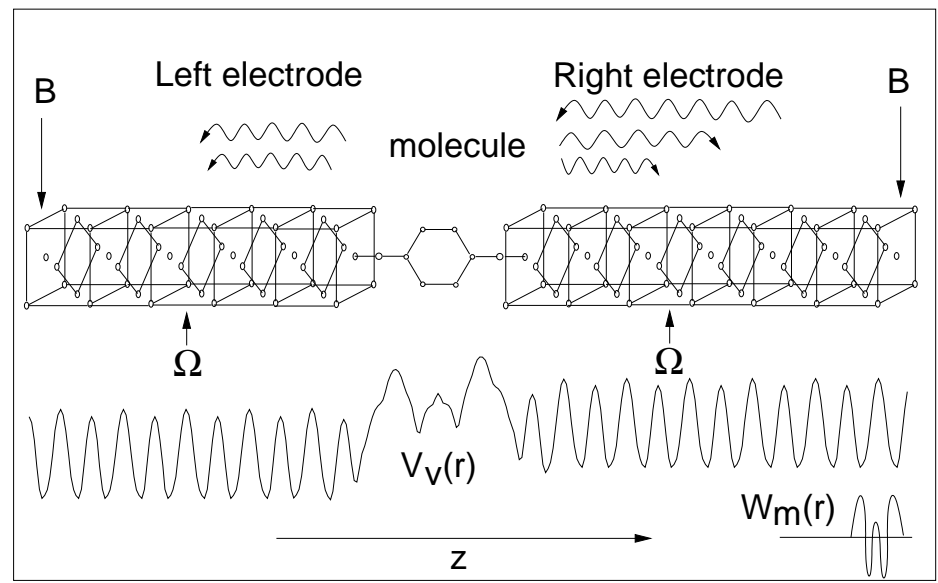
\begin{figure}[floatfix]
\caption{
A schematic view of the calculated system. }
\label{Fig1}
\end{figure}

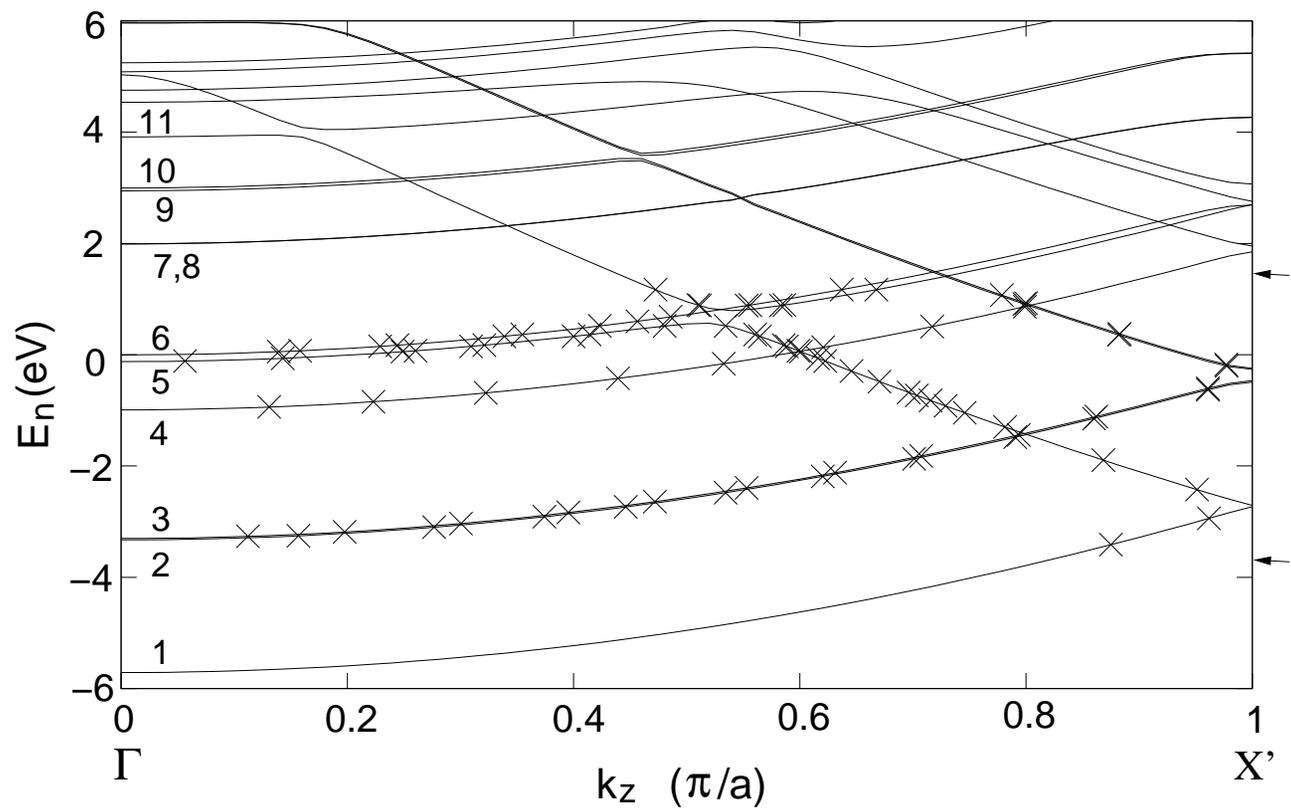
\begin{figure}[floatfix]
\caption{
The band structure of the electrode. Each continuous line from 
$\Gamma$ to $X'$ is denoted as one band. The zero is the electrode 
Fermi energy. The crosses are the $k_n^R$ points, see text for details.}
\label{Fig2}
\end{figure}

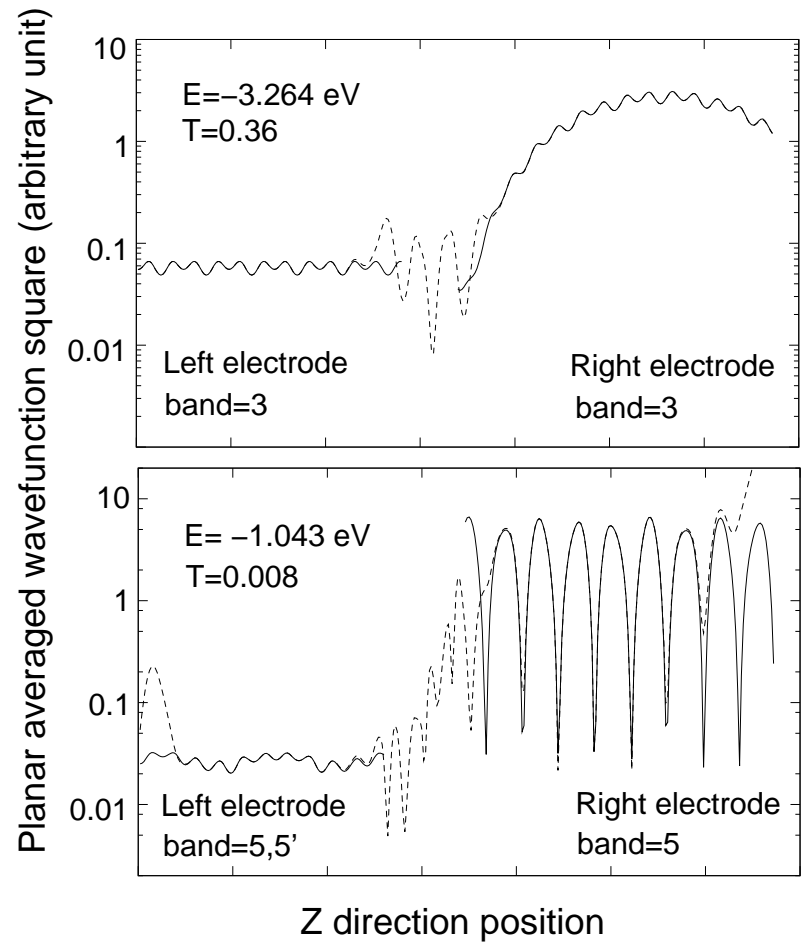
\begin{figure}[floatfix]
\caption{
The constructed scattering states from 
Eq(4). The dashed and solid lines correspond to 
the first and second lines of Eq(4) respectively. 
T is the transmission coefficient, E is 
the eigen energy, and band number indicate the $n$
of $\phi_n$ in Eq(4). The bias of the system is 1V. }
\label{Fig3}
\end{figure}

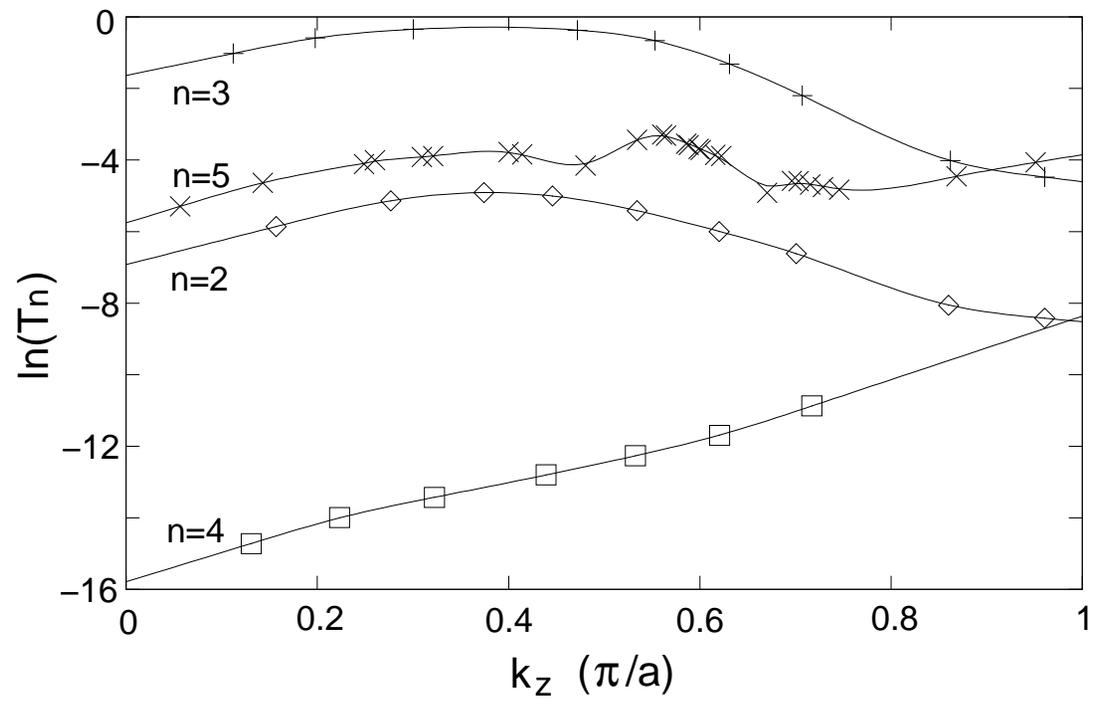
\begin{figure}[floatfix]
\caption{
The calculated transmission coefficients $T_n(k_n^R)$ (symbols)
and fitted smooth curves $f_n(k)$ (lines) for different 
bands of the electrode. The bias of the system is 1V.}
\label{Fig4}
\end{figure}

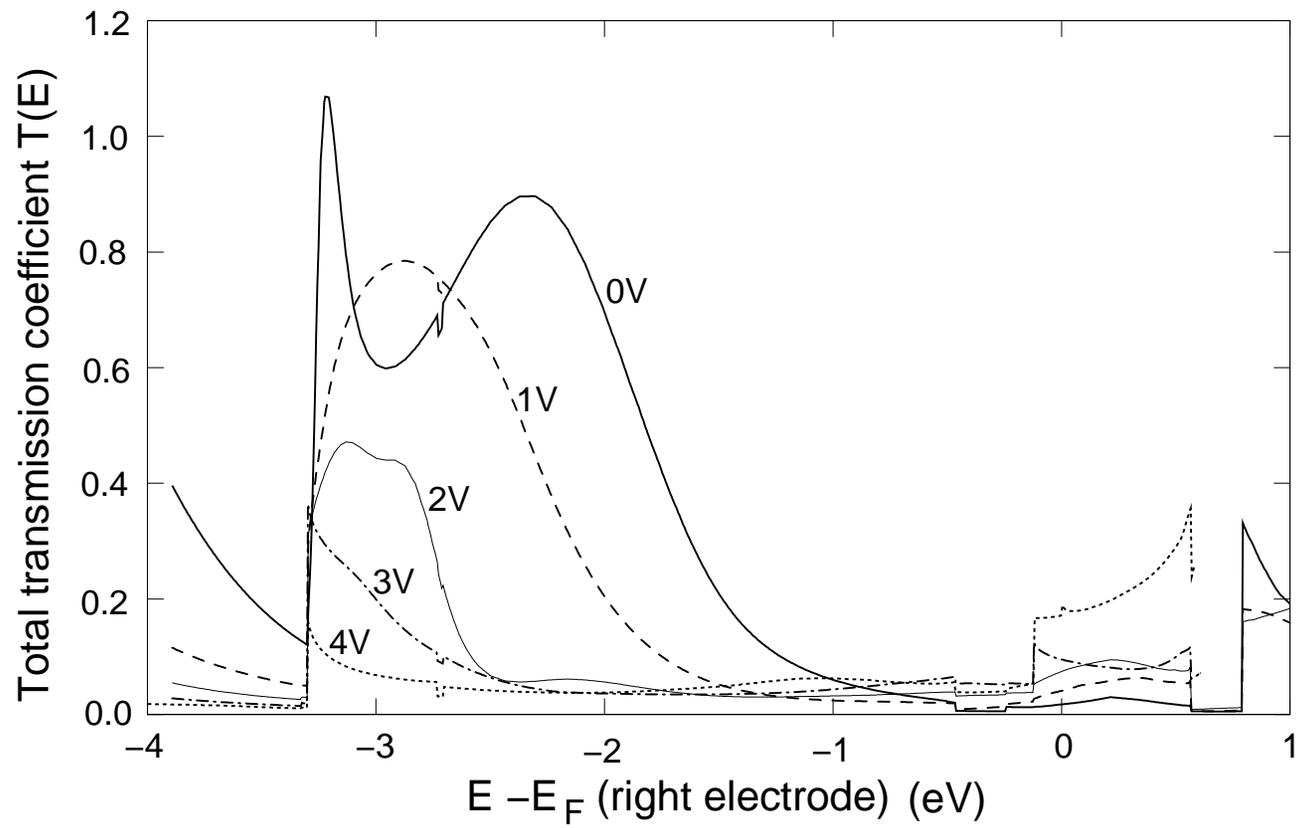
\begin{figure}[floatfix]
\caption{
The calculated total transmission coefficients $T(E)$ (which can be 
larger than 1) for different biases. The zero is the right electrode
Fermi energy. For a given bias V, there are net right to left current flow
only within the $[-V,0]$ energy window.}
\label{Fig5}
\end{figure}

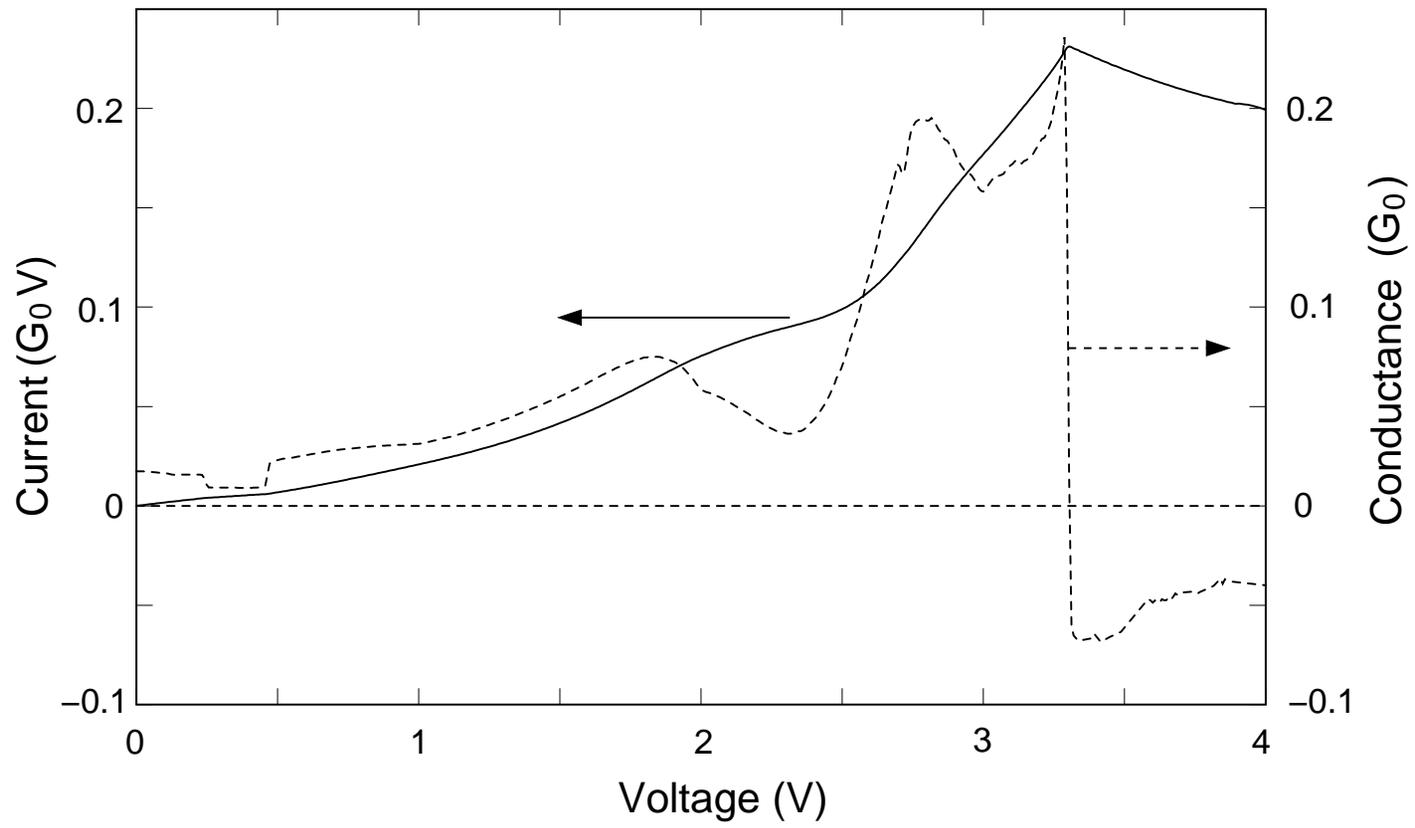
\begin{figure}[floatfix]
\caption{
The calculated I-V curve and the corresponding conductance.
$G_0=2e/h=77\mu s$. } 
\label{Fig6}
\end{figure}


\begin{thebibliography} {1}

\bibitem{Greens} J. Taylor, H. Guo, J. Wang, Phys. Rev. B {\bf 63}, 245407 (2001); 
P. Damle, A. Ghosh, S. Datta,  Phys. Rev. B {\bf 64}, 201403 (2001);
M. B. Nardelli, J.-L. Fattebert, J. Bernholc, Phys. Rev. B {\bf 64}, 245423 (2001);
M. Brandbyge, {\it et.al}, Phys. Rev. B {\bf 65}, 165401 (2002);
E. Louis, {\it et.al}, Phys. Rev. B {\bf 67}, 155321 (2003).  

\bibitem{transmatrix1} 
K. Hiros, M. Tsukada, Phys. Rev. B {\bf 51}, 5278 (1994); 

\bibitem{transmatrix2} 
H.J. Choi, J. Ihm, Phys. Rev. B {\bf 59}, 2267 (1999).

\bibitem{transmatrix3}
D. Z.-Y. Ting, E. T. Yu, T.C. McGillha, Phys. Rev. B {\bf 45}, 3583 (1992).  

\bibitem{DiVentra}
M. Di Ventra, S.T. Pentelides, N. D. Lang, Phys. Rev. Lett. {\bf 84}, 979 (2000);
M. Di Ventra, N. D. Lang, Phys. Rev. B {\bf 65}, 45402 (2001). 

\bibitem{Reed} M.A. Reed, {\it et.al}, Science {\bf 278}, 252 (1997). 

\bibitem{Roland}
C. Roland, V. Meunier, B. Larade, H. Guo, Phys. Rev. B {\bf 66}, 35332 (2002). 

\bibitem{PEtot}
http://crd.lbl.gov/~linwang/PEtot/PEtot.html

\bibitem{footnote}
When the energy $E_i$ is close to an extremum of a band energy (e.g, at the $\Gamma$ 
and $X'$ points, or inside an anticrossing band gap of Fig.2), an evanescent state
might exist at the middle electrode cell $\Omega$ where the decomposition is 
calculated. These evanescent states come from the complex $k$-point band structure 
of the electrode [see: Y.C. Chang, Phys. Rev. B {\bf 25}, 605 (1982)]. If the
calculated electrode is very long, these evanescent states should have decayed 
to zero in the middle of the electrode. But in practice, and especially when 
the energy is close to the extrema, these evanescent states can have significant
contributions. We have used $u_n(k_m)exp(ik_m z)$ at those extrema $k_m$ to 
approximate the evanescent states, and include them as $\phi_n(r)$  in the decomposition of 
$\phi_{i,(l)}$. Note that, these evanescent states carry no currents since
$d E_n(k)/d k|_{k=k_m}=0$. 

\bibitem{PAM}
N.V. Smith, N.B. Brookes, Y. Chang, and P.D. Johnson, Phys. Rev. B
{\bf 49}, 332 (1994). 

\end{thebibliography}
\end{document}